
\input phyzzx
\PHYSREV
\hbox to 6.5truein{\hfill UPR-0517T}
\hbox to 6.5truein{\hfill July, 1992}
\title{FINAL-STATE-INTERACTION SIMULATION OF $T$-VIOLATION
IN THE TOP-QUARK SEMILEPTONIC DECAY}
\author{Jiang Liu}
\address{Department of Physics, University of Pennsylvania, Philadelphia,
         PA 19104}
\abstract

The standard electroweak final-state interaction induces
a false $T$-odd correlation in the top-quark semileptonic
decay.  The correlation parameter
is calculated in the standard model and found to be
considerably
larger than those  that could be produced by
genuine $T$-violation effects in a large class of theoretical models.
$$ $$
PACS\#: 11.30.Er, 13.30.Ce.
\endpage
\chapter{Introduction}

Final-state interactions play an important role in the determination
of $CP$ and $T$ violation.  A test for $CP$ violation
is to compare the partial decay rates of a particle and its antiparticle.
In this case final-state interactions are necessary since in their
absence the partial decay rates are equal from $CPT$ invariance even if
$CP$ is violated.  General formalism for calculating such partial rate
differences based on $CPT$ invariance and unitarity has recently been
developed\rlap,\Ref\Lincoln{
      L. Wolfenstein, Phys. Rev. D43, 151 (1991).}
and its applications to $B$ meson decays (Ref. \Lincoln) and
to  $t$-quark decays\Ref\Soares{
      J. M. Soares, Phys. Rev. Lett. 68, 2102 (1992).}
have revealed some interesting relations between
final-state interaction and $CP$ violation observables in weak decays.

A test\Ref\lwolf{For a review see:  L. Wolfenstein, Ann. Rev. Nucl.
Part. Sci. 36, 137 (1986).}
 for $T$ violation is to observe a ``$T$-odd correlation'',
such as those of the form $\vec{\sigma}\cdot(\vec{p}_1\times
\vec{p}_2)$ where $\vec{\sigma}$ is a spin and $\vec{p}_1$ and
$\vec{p}_2$ are  momenta.  In contrast to the partial
decay difference, a $T$-odd correlation can be produced
by final-state interactions even if $T$ invariance holds.
Thus, to use such correlations as a test of $T$ violation
the final-state-interaction effect must be negligible or calculable.

This paper will be concerned with the $t$-quark semileptonic
decay $t\to bW\to b\nu_{\ell}\bar{\ell}$ in the standard model.
Copious production of $t$-quarks
at future high-energy colliders such as the SSC and the LHC
have aroused considerable interest in exploring the origin
of $CP$ and $T$ violation via $t$-quark interactions\rlap.\Ref\Eilam{
      G. Eilam, J. L. Hewett, and A. Soni, Phys. Rev. Lett. 67,
      1979 (1991);
      {\it{ibid}}, 68, 2103 (1992); G. L. Kane, G. A. Ladinsky,
      and C.-P. Yuan, Phys. Rev. D45, 124 (1992);
      D. Atwood and A. Soni, Phys. Rev. D45, 2405 (1992);
      \splitout
      W. Bernreuther, T. Schr\"{o}der, and T. N. Pham, Heidelberg
      report, HD-THEP-91-39 (1992);
      A. Soni and R. M. Xu, Phys. Rev. Lett. 69, 33 (1992);
       C. R. Schmidt and M. E. Peskin,
      Phys. Rev. Lett. 69,  410 (1992);\splitout
      R. Cruz, B. Grz\c{a}dkowski and J. F. Gunion, UC Davis report,
      UCD-92-15 (1992).}
In particular, a recent study\Ref\Gunion{
      B. Grz\c{a}dkowski and J. F. Gunion, UC Davis report, UCD-92-7
      (1992) (to appear in Phys. Lett. B.).}  of the
possibility of using the $T$-odd correlation has shown that it
has a reasonable sensitivity to some non-standard sources
of $T$ Violation.  Since such correlations can be produced by
standard model physics alone, it is timely to undertake
a computation of the final-state-interaction effect
due entirely to the standard electroweak interaction, which,  up to the
one-loop level,  respects $T$ and $CP$ invariance in Cabibbo-allowed
weak decays such as $t\to bW^+\to b\nu_{\ell}\bar{\ell}$.

\bigskip
\chapter{Final-State-Interaction Effect}

The computation of final-state-interaction effects on the
$T$-odd correlation has long been of interest.  Early examples
of the calculation involved nuclear $\beta$ decay\rlap,\Ref\nuclear{
      C. G. Callan and S. B. Treiman, Phys. Rev. 162, 1494 (1967);
      J. C. Brodine, Phys. Rev. D1, 100 (1970).}
hyperon semileptonic decay\rlap,\Ref\hyper{
      L. B. Okun and I. B. Khriplovich, Yad. Fiz. 6, 1265 (1967)
      [Sov. J. Nucl. Phys. 6, 919 (1968)].}
and $K^{\pm,0}_{\ell 3}$ decays\rlap.\Ref\Kthree{
      N. Byers, S. W. MacDowell, and C. N. Yang, `{\it{
      High Energy Physics and Elementary Particles}},'
      IAEA, Vienna, p953 (1965);
      L. B. Okun, JETP letters, 6, 272 (1967);\splitout
       I. B. Khriplovich and L. B. Okun, Phys. Lett. 24B, 672 (1967);
       \splitout
       L. B. Okun and I. B. Khriplovich, Yad, Fiz, 6, 821 (1967)
      [Sov. J. Nucl. Phys. 6, 821 (1967)].
       J. S. Bell and R. K. P. Zia, Nucl. Phys. B17, 388 (1970);
       J. C. Brodine, Nucl. Phys. B30, 545 (1971);
       E. S. Ginsberg and J. Smith, Phys. Rev. D8, 3887 (1973);
       \splitout
       A. R. Zhitinitski\v{i}, Yad. Fiz. 31, 1024 (1980)
       [Sov. J. Nucl. Phys. 31. 529 (1980)].}
The parameter of interest is the coefficient  of the $T$-odd
correlation term  in the decay spectrum,
which in  nuclear $\beta$ decay, for instance,
has the following form in the leading approximation
$${d\Gamma\over d\Omega_e d\Omega_{\nu_e}dE_e}\sim
1+a{\vec{p}_e\cdot\vec{p}_{\nu_e}\over E_eE_{\nu_e}}
+\vec{\sigma}\cdot\Bigl[A{\vec{p}_e\over E_e}+B{\vec{p}_{\nu_e}\over
E_{\nu_e}}+D{\vec{p}_e\times\vec{p}_{\nu_e}\over E_eE_{\nu_e}}\Bigr],
\eqno(1)$$
where $\vec{\sigma}$ is the polarization of the parent nucleus and
$\vec{p}_e(E_e)$ and $\vec{p}_{\nu_e}(E_{\nu_e})$
are the electron and  neutrino momentum (energy), respectively.
In this example, the
dominant contribution arises from  electromagnetic
final-state interaction. The effect depends, among other things, on
the recoil of the decaying particle, and thus  the size of the
$T$-odd correlation parameter
$D$ is of order $D\sim \alpha E_e/M\ (Z\alpha E_e/M)$
in neutron  (nuclear) $\beta$ decay,
where $M$ is a nucleon mass.  Since $E_e$ is typically of order
$1\ MeV$,  the recoil effect, which is characterized
by the  ratio  $E_e/M$, is rather tiny.  Hence  $D$ is highly
suppressed in neutron $\beta$ decay
with $D$ typically of the order of
$10^{-5}-10^{-6}$. A considerably larger result ($10^{-3}-10^{-4}$)
can be obtained in some nuclear $\beta$ decays due to the
enhancement $Z\gg 1$
(Ref. \nuclear).   The typical value of the $T$-odd correlation
is between $10^{-3}$ and $10^{-4}$ in a neutral
$K^0_{\ell 3}$ decay.  The result in a charged $K^{\pm}_{\ell 3}$
decay is still  smaller ($10^{-5}-10^{-6}$), because there
the final-state pion is neutral and the effect can only arise from
two-loop graphs.

In terms of weak-current interactions the $t$-quark semileptonic decay
is analogous in many respects to  nuclear $\beta$ decay.  However,
the disparity between $m_t$ and $m_b$ implies that
the $T$-odd correlation in the decay $t\to b\nu_{\ell}
\bar{\ell}$
does not have a recoil suppression. Indeed, compared to  nuclear
$\beta$ decay, where the recoil effect is of order $10^{-3}$,
in the $t$ semileptonic decay such effects are given by $E_{\bar{e}}
/m_t$,
which is of order unity.  As a consequence, we
expect that the final-state-interaction contribution to the
$T$-odd correlation parameter is roughly
$$D(t\to bW\to b\nu_e\bar{e})\sim \alpha \vert Q_d\vert
{E_{\bar{e}}\over m_t}\sim
{\alpha\over 9}\sim 10^{-3},\eqno(2)$$
where $Q_d=-1/3$ is the $b$-quark charge, and
we have taken $E_{\bar{e}}/m_t\sim 1/3$.

 In what follows we will
concentrate on the decay $t\to bW\to b\nu_e\bar{e}$.  Insofar
as the lepton mass can be ignored, our result holds for the other
$t$-quark semileptonic decays as well.

A large $m_t$ implies that the decay $t\to b\nu_e\bar{e}$
proceeds dominantly through the $W$ resonance.  The smallness
of the $W$ width ($\Gamma_W/M_W\approx 0.026$) then makes the
calculation of the leading final-state-interaction effect very simple.
Neglecting the $b$-quark and  lepton masses, the leading
contributions are generated by graphs displayed in Fig. 1
with
$$M(t\to b\nu_e\bar{e})=\Bigl({ig\over\sqrt 2}\Bigr)^2
{[\bar{u}_{\nu_e}(p_{\nu_e})\gamma_{\lambda}Lv_{\bar{e}}(p_{\bar{e}})]
[\bar{u}_b(p')\Gamma^{\lambda}u_t(p)]\over
k^2-M_W^2+i\Gamma_WM_W},\eqno(3)$$
where $k=p-p'$ is the momentum transfer carried by the $W$,
$L$ and $R$ are the helicity projection operators,  and
the effective vertex $\Gamma^{\lambda}$, which includes one-loop
interaction corrections from (Fig. 1b) and (Fig. 1c),
can be parameterized as
$$\Gamma^{\lambda}=F_1(k^2)\gamma^{\lambda}L
-iF_2(k^2)m_t\sigma^{\lambda\mu}k_{\mu}R,\eqno(4)$$
where $\sigma^{\lambda\mu}={i\over 2}[\gamma^{\lambda},\gamma^{\mu}]$.
Terms  of the form $\gamma^{\lambda}R$ and
$\sigma^{\lambda\mu}k_{\mu}L$ vanish in the limit $m_b=0$.
Also, the $k^{\lambda}$ term drops out for $m_e=m_{\nu_e}=0$.
While the form factor $F_1=1+O\Bigl({\alpha\over\pi}\Bigr)$
introduces a correction to the weak interaction
charge $g$,  $F_2$ gives an anomalous moment to the
$\bar{b}tW$ vertex.

In analogous to nuclear $\beta$ decay one may define a $T$-odd
correlation parameter $D$:
$${d\Gamma\over d\Omega}=
{g^4\over (2\pi)^5}{m_tE_{\nu_e}E_{\bar{e}}\over
\vert k^2-M_W^2+i\Gamma_WM_W\vert^2}
\Bigl[\Bigl(1-{k^2\over 2m_tE_{\nu_e}}\Bigr)+
D\Bigl(1-{2E_{\bar{e}}\over m_t}\Bigr)
\vec{\sigma_t}\cdot {\vec{p}_{\bar{e}}\times \vec{p}_{\nu_e}\over
E_{\bar{e}}E_{\nu_e}}\Bigr]+...\eqno(5)$$
with
$$D=m_t^2ImF_2(M_W^2)\eqno(6a)$$
evaluated at $k^2=M_W^2$.
The ellipses in Eq. (5) refer to the other terms of no interest to us
and $d\Omega=(d^3\vec{p}_{\bar{e}}/2E_{\bar{e}})(d^3\vec{p}_{\nu_e}
/2E_{\nu_e})(d^3\vec{p}'/2p'_0)$.
In reaching $(6a)$ we have taken $F_1=1$.

The final-state interaction in nuclear $\beta$ decay
takes place between the daughter nucleus and the electron.  By
contrast, the dominant
effect in the decay $t\to bW^+\to b\bar{e}\nu_e$ arises from
$bW\to bW$ rescattering. By employing the unitarity formula given
by Wolfenstein (Ref. \Lincoln\ ) one can show that
the relevant interactions
are those which scatter a $bW^+$ state to other $bW^+$ states
with  different spin configurations.  As a result, the $T$-odd
correlation parameter is directly proportional to the absorptive part
of the form factor
 $F_2$ which connects hadron states with different helicities.
We find (the detail of the calculation is summarized in the
Appendix)
$$ImF_2(M_W^2)=-{\alpha Q_d\over 2m_t^2}\Bigl(1-{1\over 2}
{M_W^2\over m_t^2}\Bigr)+{\alpha(1+2Q_ds^2)\over 8(m_t^2-M_W^2)}
\Bigl[\Bigl({1\over c^2}-{1\over s^2}\Bigr)I_1+{2\over s^2}I_2\Bigr],
\eqno(6b)$$
where $s^2=\sin^2\theta_W$, $c^2=\cos^2\theta_W$ and
$$\eqalign{
I_1=&2+\Bigl[1+{2m_t^2M_Z^2\over (m_t^2-M_W^2)^2}\Bigr]\ln{
      M_Z^2m_t^2\over M_Z^2m_t^2+(m_t^2-M_W^2)^2},\cr
I_2=& \Bigl(1-{M_W^2\over m_t^2}\Bigr)\Bigl[1-{1\over 2}
     {M_W^2\over m_t^2}+2{M_Z^2\over m_t^2-M_W^2}+3
     {M_W^2M_Z^2\over (m_t^2-M_W^2)^2}\Bigr]\cr
&    +{M_Z^2\over m_t^2-M_W^2}\Bigl[2+2{M_Z^2\over m_t^2-M_W^2}
     +3{M_W^2M_Z^2\over (m_t^2-M_W^2)^2}\Bigr]\ln
     {M_Z^2m_t^2\over M_Z^2m_t^2+(m_t^2-M_W^2)^2}.\cr}\eqno(6c)$$
In Eq. $(6b)$ the first term comes from the photon graphs and
the second from the $Z$.  For a very heavy top
the result is dominated by the
$Z$ exchange diagram and has a logarithmic dependence on $m_t$.
Asymptoticaly it approaches
$$\lim_{m_t\to\infty}D={\alpha\over 6}\Bigl[1+{3\over 4}\Bigl(1
      -{2s^2\over 3}\Bigr)\Bigl[{2\over c^2}+\Bigl({1\over c^2}
     -{1\over s^2}\Bigr)\ln {M_Z^2\over m_t^2}\Bigr]\Bigr].
\eqno(7)$$

The numerical results for $D$ from Eqs. $(6a)$ to $(6c)$ are
summarized\Ref\rad{
      We have taken
       $\alpha=\alpha(M_Z)=1/127.9\pm 0.2$:
       A. Sirlin, Phys. Rev. D22, 971 (1980).}
in Table 1 for $m_t$ between $100\ GeV$ and $200\ GeV$.
One sees that $D$ is between $1\times 10^{-3}$ and $5\times 10^{-3}$,
as we expected from the simple dimensional argument Eq. (2).
The result  shows a slow increase
with larger values of $m_t$ in this region.

The $T$-odd correlation  may be reparameterized in
terms of an asymmetry parameter $A$, which is related to
the difference of the decay $W^+\to \bar{e}\nu_e$ occuring
in the opposite sides of
the $\vec{\sigma}_t\times \vec{p}'$ plane (Ref. \Gunion)
$$A=-{3(m_t^2-M_W^2)\over  4
     (m_t^2+2M_W^2)}{m_tM_WIm(F_1F_2^*)\over \vert F_1\vert^2}
\approx -
     {3(m_t^2-M_W^2)\over 4(m_t^2+2M_W^2)}m_tM_WImF_2^*,\eqno(9)$$
where $ImF_2^*$ is given by Eqs. $(6b)$ and $(6c)$ with an additional
over all minus sign.
The results for $A$ are summarized in the last column of Table 1.
They vary from $1\times 10^{-4}$ to $1\times 10^{-3}$ for
$m_t=100-200\ GeV$.  In comparison with
the maximal-allowed  $T$ violation  effect in the models
considered in Ref. \Gunion\  in which $A<5\times 10^{-5}\sim 5\times
 10^{-4}$,
  the standard
model final-state interaction produces a much larger false effect.

It is difficult to calculate the $T$-odd parameter to an accuracy
of $\sim 30\%$. The major
theoretical uncertainties of the present calculation come from neglecting
QCD corrections, which introduce  a sizable interference between
the absorptive part of $F_1$ from electroweak interactions
and the real part of $F_2$ from QCD.
An order of $\sim (1\sim 10)\%$ correction due to this effect alone
is possible.  A still more complicated contribution arises from the
interference between $ImF_2$ calculated above
 and the real part of $F_1$ due to QCD.
 Other uncertainties  arise from  neglecting (1)
the $WZ$ threshold effect (relevant if $m_t>M_W+M_Z+m_b$) and (2)
all the box-diagrams. The contribution of the latter also
depends on the angle between $\vec{p}_{\bar{e}}$ and $\vec{p'}$ in a
rather complicated way.
All of these contributions are suppressed by the ratio $\Gamma_W/M_W$,
however.
The calculation  of these
next-leading terms would be crutial should future experiments
approach  the precision of $D\sim 10^{-3}$.

$T$-odd correlations of the form $\vec{\sigma}_{\bar{e}}\cdot
(\vec{p}_{\nu_e}\times \vec{p}_{\bar{e}})$, $\vec{\sigma_b}\cdot(
\vec{p}_{\nu_e}\times\vec{p}_{\bar{e}}
)$ and $P$- and $CP$-odd correlation of the form
 $\vec{\sigma}_t\cdot (\vec{\sigma}_b\times \vec{p}')$
 are much more difficult to measure experimentally,
and thus will not be considered in this paper.

\bigskip

\chapter{Conclusion}

We have calculated the $T$-odd correlation $\vec{\sigma}_t\cdot
(\vec{p}_{\bar{e}}\times \vec{p}_{\nu_e})$
 induced
by the standard electroweak final-state interactions
in the decay $t\to bW^+\to b\bar{\ell}\nu_{\ell}$, and found that the
result has
a logarithmic dependence on the $t$-quark mass and is dominated
by  the  $bW\to bW$ rescattering due to a $Z$ exchange in the
heavy top limit.  For $m_t$ in the range  $100\ GeV$ to $200\ GeV$
the correlation parameter $D$ defined in Eq. (5)
 is between $1\times 10^{-3}$ to $5\times
10^{-3}$, and the asymmetry parameter $A$ given by Eq. (9) is between
$1\times10^{-4}$ to $1\times 10^{-3}$.  It is
 shown that the standard model
physics can simulate a false $T$-odd signal, with its
magnitude  exceeding  genuine $T$-violation
effects of a size that  could possibly be produced
in a large class of theoretical models.  To get rid of this pure
final-state interaction effect one may consider comparing the asymmetry
parameter for both $t\to bW^+$ and $\bar{t}\to \bar{b}W^-$,
as in the study of CP-violating  parameters
$\alpha+\bar{\alpha}$ and $\beta+\bar{\beta}$ in the
$\Lambda$ decays\rlap.\Ref\Dchang{O. E. Overseth and S. Pakvasa, Phys.
Rev. 184, 1663 (1969);
D. Chang and L. Wolfenstein,
Proc. Symp. on Intense Medium Energy Source of Strangeness,
ed. T. Goldman, New York: Am. Inst. Phys. (1983);
J. F. Donoghue and S. Pakvasa, Phys. Rev. Lett. 55, 162 (1985).}
\bigskip

\ack
I wish to thank  Paul Langacker and Lincoln Wolfenstein
 for valuable discussions and comments. I would also like to thank
Aspen Center for Physics for hospitality during the
completion of this work.
This research was supported in part by the U.S. Department of
Energy under contract DE-ACO2-76-ERO-3071 and the SSC National
Fellowship.
\endpage
\centerline{FIGURE CAPTION}

Fig. 1.  Feynman graphs generating the dominant contributions to the
$T$-odd correlation.  The calculation is carried out in the
Feynman-'t Hooft gauge.  $\phi$ is the Higgs-Goldstone-boson.

\bigskip

\centerline{TABLE CAPTION}

Table 1.  The result for the $T$-odd correlation in the $t$-quark
semileptonic decay.  The parameters $D$ and $A$ are defind in
Eq. (5) and Eq. (9), respectively.
\endpage
\centerline{APPENDIX}

We give some details of the calculation in this appendix.
The technique is standard\Ref\Tini{
    G. 't Hooft and M. Veltman, Nucl. Phys. B153, 365 (1979);
    G. Passarino and M. Veltman, Nucl. Phys. B160, 151 (1979).}
except that we use the Minkowskian metric $g^{\lambda\beta}
=diag(1,-1,-1,-1)$.  The one- and two-point functions are defined as
$$\eqalign{A(m)=&-i\mu_0^{(n-4)}\int
{d^nK\over (2\pi)^n}{1\over [K^2-m^2+i\epsilon]},\cr
B(m_1,m_2;k)=&-i\mu_0^{(n-4)}\int {d^nK\over (2\pi)^n}
  {1\over [K^2-m_1^2+i\epsilon][(K+k)^2-m_2^2+i\epsilon]},\cr}
\eqno(A.1)$$
where $\epsilon\to 0_+$,  and we use  dimensional regularization
to isolate the ultra-violet divergences.
The only relevant three-point function is
$$C_0=-i\int {d^4K\over(2\pi)^4}
{1\over[K^2-M_Z^2+i\epsilon][(K-k)^2-M_W^2+i\epsilon]
[(K+p')^2-m_b^2+i\epsilon]}.\eqno(A.2)$$
We find
$$\eqalign{&ImA(m)=0,\cr
&ImB(m_1,m_2;k)={1\over 16\pi k^2}\sqrt{\lambda(k^2,m_1^2,m_2^2)}
  \theta[k^2-(m_1+m_2)^2],\cr
&ImC_0={1\over 16\pi\sqrt{\lambda(m_t^2,M_W^2,m_b^2)}}
\ln{M_Z^2m_t^2\over M_Z^2m_t^2+\lambda(m_t^2,M_W^2,m_b^2)}
\theta[m_t^2-(M_W+m_b)^2],\cr}\eqno(A.3)$$
where $\lambda(x,y,z)=x^2+y^2+z^2-2xy-2xz-2yz$.  In evaluating
$ImC_0$ we have put all the external lines on their mass-shell.

Neglecting the $b$-quark and  lepton masses, the
final-state-interaction effect due to a photon exchange is
$$\Gamma^{\lambda}(\gamma)=-e^2Q_d4m_t
p^{\prime\lambda}R(a_1+b_1),\eqno(A.4)$$
where $a_1$ and $b_1$ are the coefficients defined in the following
integrals:
$$-i\int {d^4K\over (2\pi)^4}{K^{\lambda}\over K^2[(K-k)^2-M_W^2]
[(K+p')^2-m_b^2]}=a_1k^{\lambda}+a_2p^{\prime\lambda},\eqno(A.5)$$
$$\eqalign{-i\mu_o^{(n-4)}\int {d^nK\over (2\pi)^n}&{K^{\lambda}
K^{\beta}\over K^2[(K-k)^2-M_W^2][(K+p')^2-m_b^2]}\cr
&=b_1(k^{\lambda}p^{\prime\beta}+k^{\beta}p^{\prime\lambda})
 +b_2g^{\lambda\beta}+b_3k^{\lambda}k^{\beta}+b_4p^{\prime\lambda}
p^{\prime\beta}.\cr}\eqno(A.6)$$
We find
$$\eqalign{&a_1={B(M_W,0;k)-B(M_W,0;p)\over m_t^2-M_W^2},\cr
&          b_1=-{1\over 2}{A(M_W)-M_W^2B(0,M_W;p)\over
           m_t^2(m_t^2-M_W^2)}.\cr}\eqno(A.7)$$
It then follows from Eqs. $(A.3)$, $(A.4)$ and $(A.7)$
that the absorptive
part of $\Gamma^{\lambda}(\gamma)$ is
$$\Gamma^{\lambda}_{abs}(\gamma)=
{\alpha  Q_d\over  m_t}\Bigl(1-{M_W^2\over 2m_t^2}\Bigr)
p^{\prime\lambda}R.\eqno(A.8)$$
It can be written in a more conventional form by applying
the Gordon identity
$$[\bar{u}_b(p')p^{\prime\lambda}Ru_t(p)]=
{i\over 2}[\bar{u}_b(p')\sigma^{\lambda\mu}k_{\mu}Ru_t(p)]+....
\eqno(A.9)$$

The result due to  $Z$ exchange is
$$\Gamma^{\lambda}(Z)=-e^2(1+2Q_ds^2)m_t
p^{\prime\lambda}R\Bigl[\Bigl({1\over c^2}-{1\over s^2}\Bigr)
(-a_1'+a_2'+C_0)-{2\over s^2}(a_1'+b_1')\Bigr],\eqno(A.10)$$
where the coefficients $a_1', a_2'$ and $b_1'$ are defined analogously
$$\eqalign{-i\int {d^4K\over (2\pi)^4}&{K^{\lambda}\over
[K^2-M_Z^2][(K-k)^2-M_W^2][(K+p')^2-m_b^2]}=a_1'k^{\lambda}
+a_2'p^{\prime\lambda},\cr
-i\mu_0^{(n-4)}\int {d^nK\over (2\pi)^n}&{K^{\lambda}K^{\beta}\over
[K^2-M_Z^2][(K-k)^2-M_W^2][(K+p')^2-m_b^2] }\cr
&=b_1'(k^{\lambda}p^{\prime \beta}+k^{\beta}p^{\prime\lambda})
 +b_2'g^{\lambda\beta}+b_3'k^{\lambda}k^{\beta}
 +b_4'p^{\prime\lambda}p^{\prime\beta}.\cr}\eqno(A.11)$$
We find
$$a_1'={1\over m_t^2-M_W^2}\Bigl[B(M_Z,M_W;k)-B(M_W,0;p)-M_Z^2C_0
\Bigr],\eqno(A.12)$$
$$\eqalign{a_2'=&
-{1\over m_t^2-M_W^2}\Bigl[B(M_Z,0;p')-B(M_W,0;p)-M_Z^2C_0\Bigr]\cr
&-{2M_W^2\over (m_t^2-M_W^2)^2}\Bigl[B(M_Z,M_W;k)-B(M_W,0;p)-M_Z^2C_0
\Bigr],\cr}\eqno(A.13)$$
$$\eqalign{b_1'=-{1\over (m_t^2-M_W^2)^2}\Bigl[&
    {1\over 2}\Bigl(1-{M_W^2\over m_t^2}\Bigr)\Bigl[A(M_W)
         -M_W^2B(M_W,0;p)\Bigr]\cr
&  +{1\over2}\Bigl[A(M_Z)-M_Z^2B(M_W,M_Z;k)\Bigr]\cr
&  +2M_Z^2\Bigl[B(M_W,0;p)-B(M_Z,0;p')\Bigr]\cr
&  -3{M_W^2M_Z^2\over m_t^2-M_W^2}\Bigl[B(M_W,M_Z;k)-B(M_W,0;p)\Bigr]\cr
&  +\Bigl[2+3 {M_W^2\over m_t^2-M_W^2}+{m_t^2-M_W^2\over M_Z^2}\Bigr]
   M_Z^4C_0\Bigr].\cr}\eqno(A.14)$$
One can check that in the limit $M_Z=0$ $a_1$ and $a_1'$
become identical and so do $b_1$ and $b_1'$.
The logarithmic dependence on $m_t$ in the limit $m_t\to \infty$
arises because $\Gamma^{\lambda}(Z)$ has a term which is directly
proportional to $C_0$ (see $(A.10)$).

It then follows that the absorptive part of $\Gamma^{\lambda}(Z)$ is
$$\Gamma^{\lambda}_{abs}(Z)=-{\alpha
(1+2Q_ds^2)\over  4(m_t^2-M_W^2)}m_tp^{\prime\lambda}R\Bigl[\Bigl(
{1\over c^2}-{1\over s^2}\Bigr)I_1+{2\over s^2}I_2
\Bigr],\eqno(A.15)$$
where $I_{1,2}$ are given by Eq. $(6c)$.  Adding $(A.8)$ and $(A.15)$
we obtain the results given by Eqs. $(6a)$ to $(6c)$ of the text.
\endpage
\input tables
\begintable
\tstrut\ $ m_t\ GeV\ $:$D$&$A$\crthick
$100 $:$1.0\times 10^{-3}\ $&$\ \  1.0\times 10^{-4}\ $\nr
$110 $:$1.4\times 10^{-3}\ $&$\ \  1.7\times 10^{-4}\ $\nr
$120 $:$1.8\times 10^{-3}\ $&$\ \  2.6\times 10^{-4}\ $\nr
$130 $:$2.2\times 10^{-3}\ $&$\ \  3.6\times 10^{-4}\ $\nr
$140 $:$2.7\times 10^{-3}\ $&$\ \  4.6\times 10^{-4}\ $\nr
$150 $:$3.1\times 10^{-3}\ $&$\ \  5.7\times 10^{-4}\ $\nr
$160 $:$3.6\times 10^{-3}\ $&$\ \  6.7\times 10^{-4}\ $\nr
$170 $:$4.0\times 10^{-3}\ $&$\ \  7.6\times 10^{-4}\ $\nr
$180 $:$4.4\times 10^{-3}\ $&$\ \  8.5\times 10^{-4}\ $\nr
$190 $:$4.8\times 10^{-3}\ $&$\ \  9.2\times 10^{-4}\ $\nr
$200 $:$5.2\times 10^{-3}\ $&$\ \  1.0\times 10^{-3}\ $\endtable

$$ $$
\centerline{Table 1}
\endpage
\refout
\end